# Atomic-resolution imaging of gold species at organic liquid-solid interfaces

Sam Sullivan-Allsop[1,2]*, Nick Clark[1,2]*†, Wendong Wang[2,3]*, Rongsheng Cai[1], William Thornley[1], David G. Hopkinson[4], James G. McHugh[2,3], Ben Davies[5], Samuel Pattisson[6], Nicholas F. Dummer[6], Rui Zhang[1,2], Matthew Lindley[1], Gareth Tainton[1], Jack Harrison[1,2], Hugo De Latour[1,2], Joseph Parker[1,2], Joshua Swindell[1], Eli G. Castanon[2,3], Amy Carl[2,3], David J. Lewis[1], Natalia Martsinovich[7], Christopher S. Allen[4,8], Mohsen Danaie[4], Andrew J. Logsdail[6], Vladimir Fal'ko[2,3], Graham J. Hutchings[5], Alex Summerfield[2], Roman Gorbachev[2,3,9]†, Sarah J. Haigh[1,2]†

[1]Department of Materials, University of Manchester; Manchester M13 9PL, UK

[2]National Graphene Institute, University of Manchester; Manchester M13 9PL, UK

[3]Department of Physics and Astronomy, University of Manchester; Manchester M13 9PL, UK

[4]ePSIC, Diamond Light Source, Harwell Science and Innovation Campus; Didcot OX11 0DE, UK

[5]Max Planck−Cardiff Centre on the Fundamentals of Heterogeneous Catalysis FUNCAT, Cardiff Catalysis Institute, School of Chemistry, Cardiff University, Cardiff, CF24 4HQ, UK

[6]Cardiff Catalysis Institute, School of Chemistry, Cardiff University; Cardiff CF10 3AT, UK

[7]Chemistry, School of Mathematical and Physical Sciences, University of Sheffield; Sheffield S3 7HF, UK

[8]Department of Materials, University of Oxford; Oxford OX1 3PH, UK

[9]Henry Royce Institute for Advanced Materials, University of Manchester; Manchester M13 9PL, UK

* These authors contributed equally to this work.

† Corresponding author. e-mail:

nick.clark@manchester.ac.uk, roman@manchester.ac.uk, sarah.haigh@manchester.ac.uk

## Abstract

Understanding solid-liquid interfaces at the atomic-scale is key to improved performance of heterogeneous catalysts, electrodes and membranes. Here we combine unique specimen design, record atomic resolution in situ electron microscopy, and artificial intelligence-enabled analysis to achieve a step change in quantitative understanding of interfacial atomic behaviour. We create the first graphene liquid cells with organic solvents and employ them to track over $10^6$ gold adatoms and clusters at a graphene surface immersed in acetone and cyclohexanone. We reveal dynamic correlated behaviour of gold adatom monomers, dimers, trimers and clusters, strongly influenced by each other, the solvent properties, and the atomic lattice of the substrate, in good agreement with theoretical calculations. We use the results to interpret differences in catalytic activity towards the industrially important acetylene hydrochlorination reaction. This new capability for exploration of atomic scale chemistry could enable rational design of future catalysts, membranes and electrodes with improved functionality.

# Introduction

Atomically dispersed metal species can combine the beneficial properties of both heterogenous and homogeneous catalysts, as well as providing electrodes, membranes and sensors with exceptional performance (*1–3*). Atomic dispersal of the active species provides higher efficiency and greater selectivity with reduced metal loading compared to supported nanoparticle equivalents (*1, 2*). Gold on carbon (Au-C) single-atom catalysts (SACs) represent a particular example where isolated gold atoms provide functionality that does not exist in the equivalent nanoparticle system. Au-C SACs have shown exceptional performance for acetylene ($C_2H_2$) hydrochlorination (*4, 5*), providing a safer alternative to the mercury-based catalysts which are used commercially to produce ~13 million tonnes of vinyl chloride monomer annually, fulfilling the world's need for polyvinyl chloride (PVC) based polymers (*6*). Understanding how to tune catalyst synthesis to control the atomic dispersal and local structural configuration is key to advancing this technology (*7*) but progress has been held back by the challenge of quantitative characterisation of single atomic species, especially under realistic environmental conditions. One particular aspect that makes atomically dispersed Au catalysts promising as SACs are the recent advances in synthesis via wet impregnation using low-polarity organic solvents, replacing established mercuric or strong acid based synthesis methods (*8*). High angle annular dark field (HAADF) scanning transmission electron microscopy (STEM) observations of the synthesised catalysts in the vacuum of the TEM were used to qualitatively verify the presence of atomically dispersed metal on the support material (*4, 8*). Yet despite thorough catalytic study in a range of solvents and wet impregnation synthesis conditions, this novel system remains largely unexplored at the atomic level and no characterisation method has been able to probe the organic liquid-solid interface during synthesis.

Transmission electron microscopy is one of the few techniques able to visualise individual atomic species on a range of different supports, usually limited to a small number of 'representative' images for the solid exposed to the TEM vacuum. Combining TEM with graphene liquid cells (GLCs) makes it possible to obtain atomically resolved images of solid particles (*9*) and adatoms (*10*) at solid-liquid interfaces. However, previous GLC studies have been limited to aqueous solutions, since organic solvents are incompatible with the polymeric supports used when sealing the graphene cell using conventional transfer techniques (*11*), which prevents their being applied to study wet impregnation synthesis of Au SACs. Furthermore, uncontrolled drying of the solution occurs during the conventional graphene liquid cell fabrication, and this has been shown to increase the concentration of the encapsulated solution by up to 3 orders of magnitude (*12*).

Here, we present the first atomic resolution imaging and single adatom tracking in non-aqueous environments. We investigate the atomic dispersion of gold on thin graphite in contact with organic liquids (acetone and cyclohexanone), including the first quantification of monomer, dimer and trimer formations and the dynamics of adatom clustering at the solid-liquid interface. Our advanced TEM liquid cell fabrication methodology (*13*) enables control of solution concentration and artificial intelligence-enabled analysis of over 1,000,000 atomic counts.

# Results

To expand the capabilities of the graphene liquid cell TEM platform beyond aqueous chemistry and to achieve quantitative data for atomic species, we have developed a new nanofabrication strategy for the cells. It allows for the use of organic solvents (as well as strong acids and alkali solutions) by eliminating ubiquitous organic polymers in the cell assembly process. Instead, to support the 2D liquid cells during their assembly and STEM measurements, we employ cleaner and more chemically and thermally stable silicon nitride membranes with lithographically patterned hole arrays (**Figure 1a**). The complete cells comprise thin graphite (3 nm) windows on either side of a ~30 nm thick hexagonal boron nitride (hBN) spacer, with pre-patterned holes serving as the "wells" containing liquid, **Figure 1b**. The total cell thickness is thin enough for high resolution STEM imaging, while also allowing us to image only one thin graphite window at a time due to the limited depth-of-field. The final "filling and sealing" step was performed under immersion in a bulk quantity of the target liquid, **Figure 1a**, rather than during the drying process, which offers significantly greater control over the concentration of the encapsulated solution than previous methods (*12*). Furthermore, by eliminating all polymers from the process we remove the dominant source for hydrocarbon contamination of the cells, resulting in extremely clean surfaces and interfaces (*14*). Full details of specimen preparation can be found in Supplementary Information (SI) section 3 including verification of the presence of trapped liquid by electron energy loss spectroscopy.

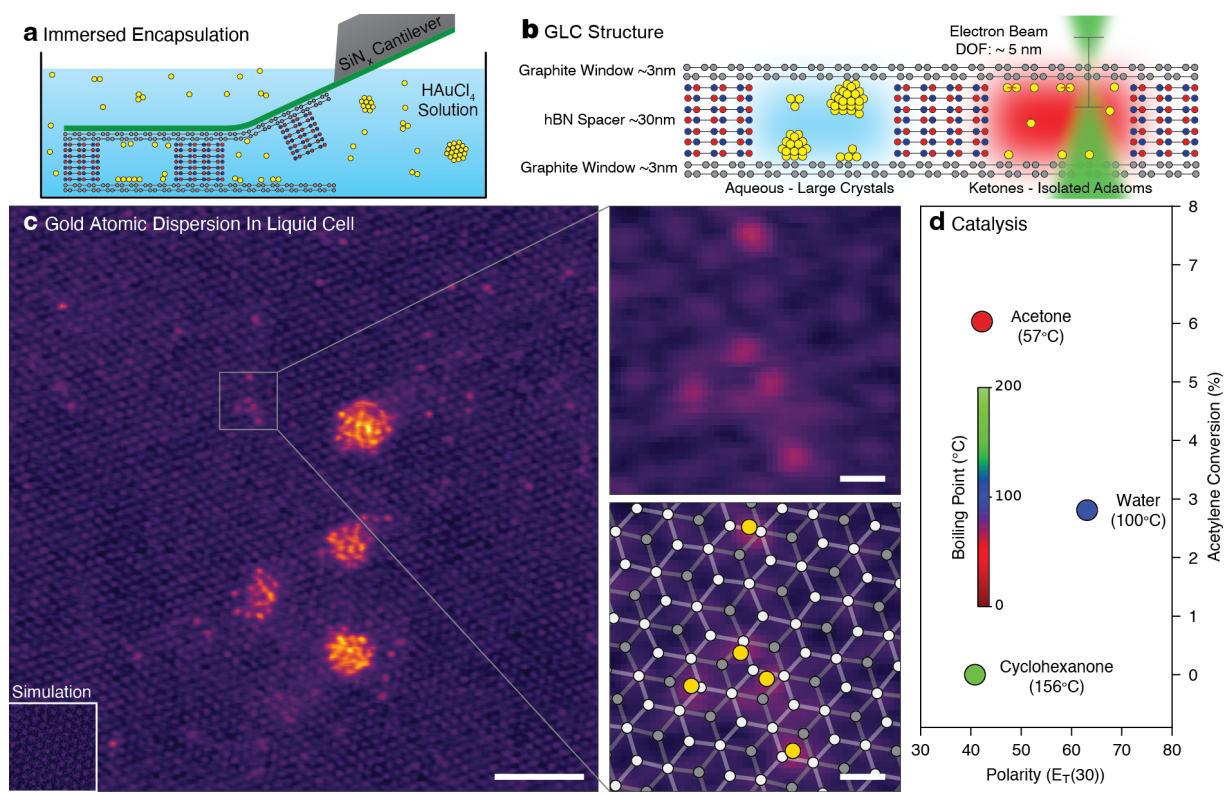

*Figure 1: Au adatoms at the graphene-solvent interface.* (a) Schematic of the fabrication of the graphene liquid cell (GLC) using a Si/SiNx cantilever. The encapsulated gold concentration is approximately equal to the bulk concentration. (b) Schematic cross section of the GLC structure. (c) Combined HAADF-STEM and BF-STEM micrographs of the final GLC containing atomically dispersed gold on the graphite lattice in contact with liquid acetone. Only one window is in focus due to the limited depth of field of the STEM probe. The inset in the bottom left of the image shows a simulated image of a graphite window at the same scale. The right top panel shows an enlarged area of the image in c highlighted by the grey square. Right lower panel is the measured positions of the gold (yellow) and the graphite lattice (alternative AB stacked graphene layers in white and grey) from the selected location. (d) Initial acetylene conversion and polarity values for Au/graphite catalysts prepared by wet impregnation in acetone, cyclohexanone and water (see SI section 6 for further details). Scalebars in (c) left: 2 nm, right: 200 pm.

**Figure 1c** shows a compound HAADF/bright field (BF) STEM image of a graphene liquid cell filled with a solution of acetone and 16 mM $HAuCl_4$. Individual Au atomic species are clearly visible on the graphite lattice as orange dots despite the liquid background and relatively weak electron signal from carbon atoms. These Au adatoms are adsorbed onto the thin graphite window and are seen to freely diffuse across it, while those in solution move too fast to be resolved (videos and image series are provided in SI section 1). Furthermore, the exceptional image fidelity allows us to map out graphite lattice sites with respect to the positions of the gold adatoms as shown for the enlarged region from, **Figure 1c**. The hexagonal pattern of the graphite lattice results from the greater intensity of carbon atoms in the top layer that have carbon atoms directly underneath (see inset simulation and SI section 2 for further details).

Along with atomically dispersed species, amorphous clusters of Au ranging from few-atom to few-nm in diameter (**Figure 1c**), were also observed in the GLC as well as several larger crystalline particles. In contrast, when the GLC is filled with water and the same 16 mM concentration of $HAuCl_4$ gold salt, no atomically dispersed Au species were observed, and the Au was only present as large crystalline nanoparticles with diameters greater than 5 nm (SI Figure S6). The clear absence of any atomically resolved Au species in the aqueous GLC explains the poor catalytic function of the Au-C catalysts synthesised with water towards acetylene hydrochlorination, since this reaction is thought to depend on the presence of isolated adatoms (*4, 5*), (**Figure 1d**). Indeed the aqueous Au-C catalyst resulted in conversion values equal to the graphite support with no Au. The atomic dispersion of the gold in acetone solvent can be attributed to the much lower polarity of acetone over water, leading to stronger Coulomb repulsion of gold ions in bulk liquid. However, this explanation raises the question of why cyclohexanone is not a similarly effective solvent for wet impregnation synthesis of Au-C SAC, having similar chemical functionality and polarity to acetone, but generating Au-C catalysts with no activity towards acetylene hydrochlorination (**Figure 1d**, see SI section 6 for further details of catalytic testing).

To quantify the differences in the distribution of Au atomic species on graphite in contact with acetone and cyclohexanone, we have compared more than 4,000 HAADF- and BF-STEM images for GLCs in both solvents. A semi-automated, AI-enabled, image analysis methodology was used, which located Au species and assigned them to clusters based on a cut-off search radius of 0.4 nm (full details in SI section 4). Example HAADF STEM images of such clusters with their assigned number of atomic species (*n*) in acetone and cyclohexanone are shown **Figure 2a**. Quantification of the cluster size distributions for acetone and cyclohexanone is shown in **Figure 2b** based on measurement of over 5,000 clusters for each solvent. We note that the actual number of atomic species in the larger clusters is likely to be greater than the assigned *n* value, as this method assumes the clusters are flat and measures only the number of 2D projected atomic sites.

The acetone and cyclohexanone atomic species follow the distribution expected for a random dispersion (black line on **Figure 2b**) up to *n*=3 atoms. Above *n*=3, the probability of finding the Au sites together in a cluster increases significantly compared to a random distribution, demonstrating the earliest stages of solid nanoparticle nucleation. Both solvents show a plateau in the probability of nanoparticle occurrence around *n*=19, corresponding to diameters of ~1.1 nm, with acetone falling off more rapidly than cyclohexanone at higher values of *n*. We therefore categorise locations with $4 \leq n \leq 18$ atomic sites as 'small amorphous clusters', $n \geq 19$ as 'large amorphous clusters', and any particles with a regular atomic arrangement as 'crystalline particles'. The latter are found for diameters > 1.1 nm and feature a distinct out-of-plane shape, while the amorphous clusters remain mostly flat (**Figure 2a**). Notably, no systematic changes in the atomic distributions or structure in the acetone or cyclohexanone GLCs were observed for samples produced and imaged after different periods of time (Figure S16), suggesting a steady state is present at the solid-liquid interface. In contrast, in the aqueous systems the crystalline particles were qualitatively observed to be larger if more time had elapsed between preparation and imaging.

**Figure 2c** shows the probability to observe an adatom in a cluster of a certain size for the different solvents. This reveals that cyclohexanone has a significantly larger number of crystalline nanoparticles, with 8% of the Au atoms in acetone solvent contributing to these compared to 72% in cyclohexanone. The remaining atoms in both acetone and cyclohexanone show a roughly equal ratio of isolated species to amorphous clusters with the mean cluster size also smaller in acetone, $\bar{n}$=12, compared to cyclohexanone, $\bar{n}$=15. Consequently, a much larger 42% of the total Au is found to be atomically dispersed for the acetone solvent, versus only 12% for cyclohexanone (for details see SI section 5.2). This observation helps to explain the dramatic difference found in their catalytic performance, **Figure 1d**.

The calculated total amount of Au for the full area imaged can also be used to estimate the concentration of Au in the liquid cell. Assuming the cell has a uniform thickness of 30 nm and most of the ions are adsorbed to one of the two thin graphite windows, we obtain a solution concentration of ~30 mM for both acetone and cyclohexanone, ~2 times higher than the bulk solutions nominal value. The difference is likely due to preferential adsorption of Au adatoms on the graphite surface in the bulk solution prior to cell closure, resulting in the graphite's surface "storing" a larger number of atoms than can be nominally contained in the GLC solvent. This simply means that the studied system is a "snapshot" of the bulk-interface equilibrium, compared to previous drying-based encapsulation techniques where the solute density was unreliable (*10, 12*).

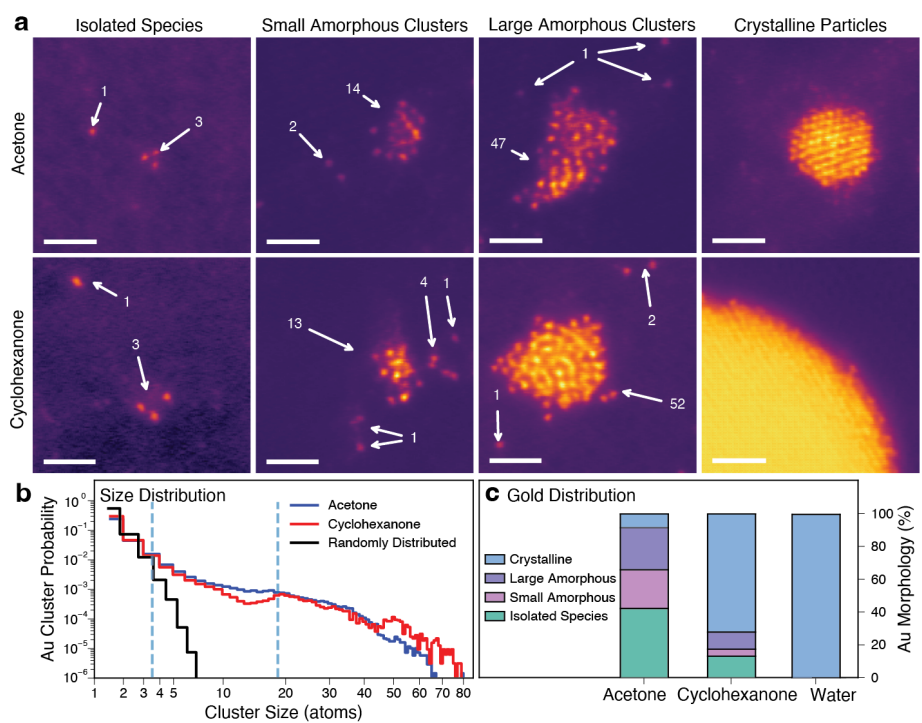

*Figure 2: Quantification of gold structures inside the GLCs.* *(a) HAADF STEM images showing the structural states of the gold within the acetone and cyclohexanone solutions (isolated species, amorphous cluster and crystalline particles). The assigned number of atomic sites for each cluster, n, visible in these example images is indicated (showing n=1,2,3,14,47 for acetone and n=1,2,3,4,13,52). (b) Size distributions of the gold species and their corresponding areal density. The vertical lines at n=3 & 18 correspond to the separation of the categories of isolated species, small amorphous and large amorphous clusters. (c) Proportions of the total measured gold existing in each structural state for acetone, cyclohexanone and water. Scalebars in (a): 1 nm.*

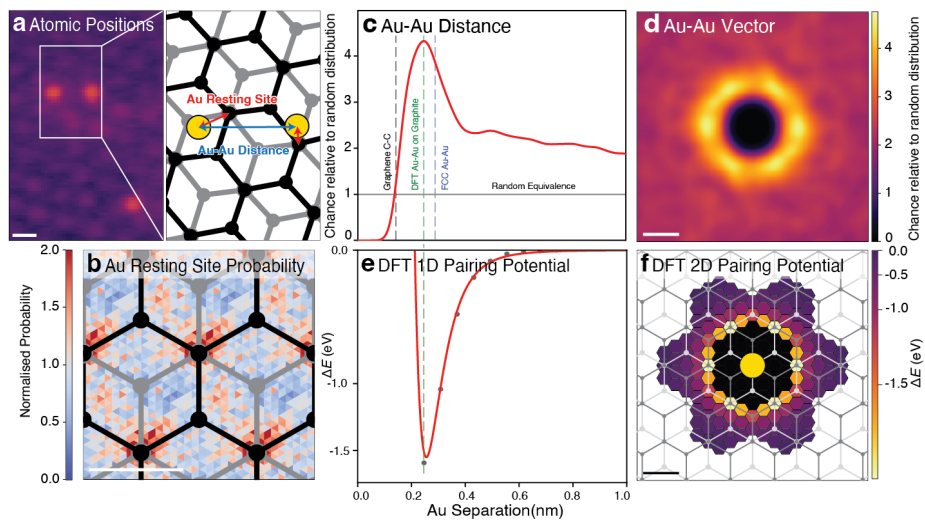

*Figure 3: Gold adatom-adatom-substrate interactions.* (a) Left: a cropped region of a combined HAADF/BF STEM image. Right: schematic of the atomic positions, and the Au resting site and the Au-Au distances for the location highlighted by the white rectangle (the graphite sublattices are shown in black and grey, and the gold adatoms in yellow). (b) The Au adatom resting site probability on the graphite lattice within an acetone environment, with the lattice stucture overlaid. (c) Au-Au 1D pair distribution function (PDF) relative to a random distribution for the isolated Au species in acetone filled GLCs. The graphene carbon-carbon separation, DFT predicted stable Au-Au distance on graphite(15) and the the FCC crystal Au-Au bond length (16) are plotted as dashed lines for reference. Inset shows the PDF at larger Au-Au separation distances. (d) Au-Au 2D PDF relative to the random distribution for the isolated Au species in acetone liquid filled GLCs. The isolated species are characterised by n= 1,2 or 3. (e) DFT calculation of potential energy for dimer as a function of Au-Au separation on graphite, with one gold atom fixed at the $A_1$ site and with the other moving along the zig-zag direction (c & e have a shared x-axis). (f) The DFT potential energies for the second gold atom placement in different sites on the graphite lattice, when the first gold atom is in the most energetically favourable $A_1$ position. The graphene lattices are overlaid in light and dark grey, and the gold atom in yellow. All the scalebars are 200 pm.

We now turn our attention to analysis of the gold adatom resting site preferences on the graphite surface (**Figure 3a**). Our density functional theory (DFT) analysis as well as earlier studies indicate that four preferential adhesion sites exist for gold on graphite in vacuum (*15*, *17*). These are: $A_1$ – atop the carbon atom for the sublattice where there is another carbon immediately beneath it; $A_2$ – atop the carbon atom where there is no carbon in the layer directly beneath it; B (bridge) – in the middle of the C-C bond and H (hollow) – in the centre of the carbon hexagon. By comparing 3,365 gold adatom positions (determined from HAADF images) to the nearest identified carbon column position (from the simultaneously acquired BF images) we plot two-dimensional histogram showing the experimental probability of finding a gold adatom at a specific resting location relative to graphene's unit cell (**Figure 3b**). We find that in both acetone and cyclohexanone there is a strong preference to occupy the $A_1$ site with all other locations being equally less likely to occur. This experimental observation agrees with our DFT calculations which predict $A_1$ is the most favourable resting site, but disagrees with an earlier DFT study that predicted both $A_1$ and $A_2$ to have similar binding energies (*15*). The difference can be attributed to the absence of solvent molecules in the DFT problems solved.

Interestingly, we observe that isolated gold species exhibit strong collective behaviour - almost half of all the Au adatoms in acetone are in dimers or trimers. These metastable formations are seen for extended periods of time (minutes), diffusing in the flat configuration across the graphite surface. This has important implications for catalytic behaviour as dimers have been recently suggested to be particularly effective "correlated" SACs (*18*). We analyse adatom-adatom distances using 1D radially averaged pair distribution functions (normalised by a simulated random distribution) in **Figure 3c** (see SI section 4 for further details). A strong preference exists for pairing with a Au-Au distance of 0.25 nm, matching the next-nearest-neighbour distance of the graphite lattice (peak in **Figure 3c** and brightest ring feature in **Figure 3d**). This generally agrees with our DFT calculations in vacuum, which show large, ~2 eV, favourable gold dimer formation energies arising from both gold-gold and gold-graphite interactions (see SI Section 7). In our model, Au-Au pair formation energy peaks when both gold atoms are situated in adjacent $A_1$-$A_1$ sites, with second most preferred being the $A_1$-$A_2$ configuration before it decays according to ~$r^{-3}$ with the dimer length $r$ (**Figure 3e**). Importantly, we observe negligible charge transfer between gold adatoms and graphite, indicating that most likely the gold ions present in bulk liquid become neutral upon adsorption onto the graphite surface. This is also confirmed by the remarkable stability of the dimers under electron beam irradiation, which would separate by Coulomb repulsion if the gold adatoms carried a significant charge (see SI section 5.3 &10 for further consideration of electron beam effects).

We further analyse the directionality of the observed dimers using the 2D pair distribution function aligned relative to the underlying graphene lattice. A 6-fold rotational symmetry is observed at the $A_1$-$A_1$ configurations, with a noticeable azimuthal spread towards the adjacent $A_1$-$A_2$ and $A_1$-H configurations (**Figure 3d**), in good agreement with DFT-calculated 2D dimer formation energy map, **Figure 3f.**

The amorphous clusters show the same peak Au-Au interatomic distance as the isolated adatoms (0.25 nm) confirming that their interaction is more similar to that of isolated species (see SI Figure S20), rather than crystalline face centred cubic nanoparticles, which have larger peak interatomic distances (0.29 nm for imaging along the [111] plane) (*16*). Additional peaks are observed at larger interatomic distances approximately corresponding to the later nearest neighbours. The influence of the graphite support is less apparent in the angular distribution of the cluster data, likely due to the competing influence from the larger number of Au adatoms.

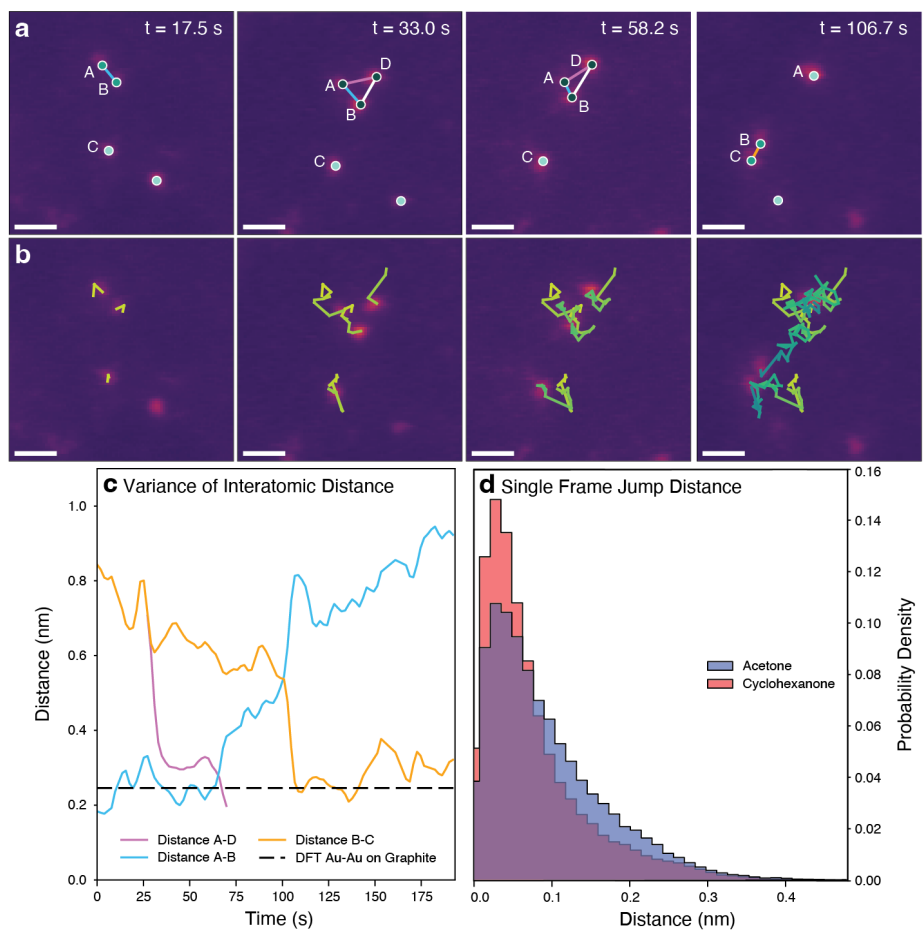

*Figure 4: Gold adatom dynamic interactions. (a & b) HAADF-STEM images from a video sequence of the gold adatoms in acetone at times of 17.5, 33.0, 58.2 & 106.7 s. (a) Highlighted locations of the Au adatoms, showing the transitions between monomer, dimer and trimer states. (b) Highlighted traces of the movement of the gold adatoms over the time period shown (coloured yellow to blue with increasing time). (c) The interatomic distances from (a) shown over the entire video sequence, AB (blue), BC (orange) and AD (purple). The black dashed line indicates the DFT Au-Au on graphite bond distance (15). (d) Single frame adatom displacement histogram for Au in liquid GLC environments, acetone and cyclohexanone. All the scalebars are 400 pm.*

The behaviour of the ions, dimers, trimers and larger clusters in the liquid cells is highly dynamic. To that end, we acquired video sequences of ~100 HAADF/BF STEM images (~1.8 s per frame, see supplementary video S1 and S2) and linked the nearest neighbours into atomic trajectories (*19, 20*) (further details in SI section 4). The Au adatoms diffuse on the graphite surface, transitioning between monomer, dimer and trimer configurations interchangeably, as can be seen in **Figure 4a,b**. Adatoms also disappear from the field of view, either due to desorption into the solution or migration outside the imaged area in a single frame. Analysis of the interatomic distances as a function of time (**Figure 4c**) highlights the dynamic nature of the process, although the preferred Au-Au spacing on graphite of 0.25 nm is again visible as the baseline interatomic distance (dashed line in **Figure 4c**). Both acetone and cyclohexanone show dynamic interplay between different atomic configurations in the isolated adatom species yet there are no structural changes in the gold that develop over time during the imaging (SI section 5.4).

Quantification of frame-to-frame jump distances reveals the distribution of frame-to-frame displacements in acetone is broader than the cyclohexanone, with larger jump distances more common in acetone (**Figure 4d**). Atomistic simulations have revealed that both acetone and cyclohexanone molecules adsorb onto the graphene surface, with the first solvation shell of both solvents oriented parallel to the surface (*21*). Our simulations have shown that for both cyclohexanone and acetone the adsorption of the solvent molecules is preferable to the adsorption of Au species, but the adsorption of the Au-solvent pairs is even more favourable. Therefore, we can expect isolated adsorbed Au species and small clusters to be capped by solvent, stabilising them and preventing large cluster growth. When the organic solvents are replaced by water, our calculations show Au is more likely to adsorb on graphene than water, therefore formation of larger gold clusters is expected to be favourable, which is all consistent with the observed formation of Au crystals in water filled liquid cells (see SI section 7 for full details of our DFT calculations).

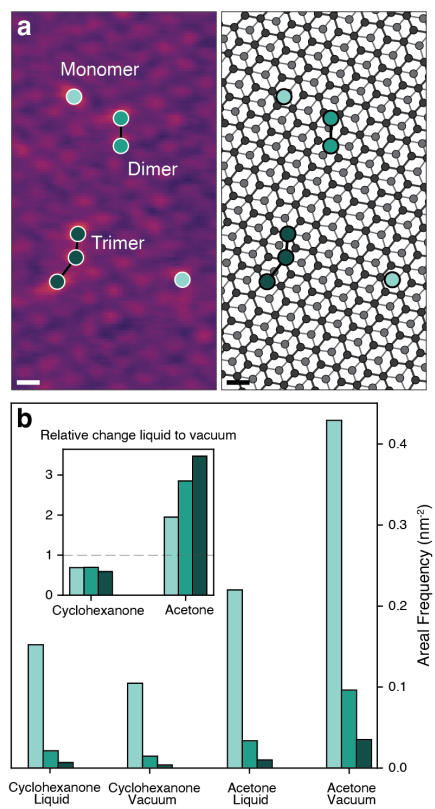

*Figure 5: Atomic dispersion of Au on graphite with and without solvent.* (a) Combined HAADF & BF STEM image (left) and corresponding schematic (right) showing the measured graphene lattice and Au adatom positions. Examples of monomer, dimer and trimer configurations are indicated. (b) Areal Frequency (nm$^{-2}$) of monomer, dimer and trimer gold adatom formations on the graphite surface in the cyclohexanone and acetone liquid cells and for the graphite support after drying (measured by conventional ex situ imaging in the TEM vacuum). The inset shows the relative change liquid to vacuum for both solvents. Scalebars in (a) are 200 pm.

For heterogeneous catalyst synthesis using wet chemical techniques, it is important to consider both the liquid phase adhesion step and the drying step for determining the final structure of the catalytic material. **Figure 5** compares the relative areal densities for isolated monomer, dimer and trimer adatoms in the acetone and cyclohexanone liquid cells with those for dried controls. Dried controls were prepared using the GLC approach but without a top graphene window and dried in air under ambient conditions. Statistically relevant measurements were obtained from 1000s of ex situ HAADF/BF STEM images of the catalyst, analysing over 1,000,000 isolated gold adatom positions (further details in SI section 4). This number of measurements is notable since even conventional ex situ TEM imaging (in vacuum) of SAC atomic assemblies typically only considers 10s of atomic sites (*1–4*) and with the previous largest ex situ analysis having considered up to 20,000 adatom positions (*22*).

**Figure 5b** shows that although the acetone liquid cells have a proportionally higher areal density of all atomically dispersed species than cyclohexanone liquid cells, both have similar percentages of monomer: dimer: trimer adatoms, with around 82±4% of isolated adatoms being monomers, 13±3% dimers and 4±1% trimers. Approximately 30% more dimers than trimers are identified on a simple counting basis, meaning that roughly equal numbers of adatoms are present in both configurations (since trimers contain 3 adatoms and dimers only 2 adatoms), likely due to the correlated adatom motion discussed in **Figure 3**.

After drying, the areal density of isolated gold adatoms has almost doubled for the acetone solvent compared to in the GLC. Monomer species increased by a factor of 2, dimers by a factor of 3 and trimers by a factor of 4, with the larger relative increase in dimers and trimers assigned simply to the greater density of atomic species on the support. In contrast, for the dried cyclohexanone sample the numbers of atomically dispersed species decreased, indicating the Au content is instead forming larger Au clusters and crystalline particles. There was no measurable change in the configuration for the clusters after drying, with the same 0.25 nm peak preferred Au-Au interatomic distance observed for dried acetone and for wet cyclohexanone. Overall, there is a 5-fold difference in the areal density of all atomically dispersed species after drying for the acetone compared to the cyclohexanone (0.12 atoms $nm^{-2}$ for the dried cyclohexanone versus 0.56 atoms $nm^{-2}$ for the acetone). We observe that the additional Au for the dried cyclohexanone is contained in large particles formed by the coffee ring effect at the edge of liquid droplets during the drying process (*23*, *24*) (see SI Figure S29). The dry samples retain their atomic dispersion when stored at ambient conditions for >6 months, likely due to residual solvents stabilising the isolated Au (see SI Figure S28). As the binding affinities of the Au(I) to acetone and cyclohexanone are similar, it is the drying behaviour that causes the large difference in the volume fraction of atomically dispersed sites. Acetone has a much lower boiling point and lower surface tension than cyclohexanone, resulting in rapid drying which preserves the favourable atomic dispersion of the organic liquid-solid interface. Drying the cyclohexanone at higher temperatures can suppress the coffee ring effect but has not improved the catalytic activity compared to acetone, likely as the increased temperature also provides energy to induce nanoparticle nucleation and growth (*25*).

In conclusion, our novel graphene liquid cell fabrication method enables the first atomic resolution studies of the solid-liquid interface for non-aqueous solvents, with the ability to encapsulate a known volume and concentration of liquid. Coupled with the liquid cells' excellent electron beam stability, we show statistically significant quantification of the resulting metal configurations and dynamics at the solvent-graphene interface, using over 1 million identified adatom locations. Importantly, we demonstrate that complex correlative behaviour can emerge from interactions between adatoms, their substrate and the surrounding environment. Our results highlight the importance of the solvent environment and deposition method on the atomic configurations of dispersed metal adatoms, and ultimately upon their catalytic stability and activity. Such quantitative in situ results can help identify the interdependence between atomic structure and catalytic activity in novel systems, with results demonstrating the importance of solvent attributes, support and drying conditions in retaining atomic dispersion for single atom catalysts. Furthermore, the ability to gain representative atomic resolution, dynamic structural characterization for known solute concentrations and non-aqueous solvents provides a powerful new tool for developing materials towards critical applications including homogeneous and heterogeneous catalysis, healthcare and clean energy systems.

**Funding:** The authors acknowledge funding from the European Research Council (ERC) under the European Union's Horizon 2020 research and innovation programme (Grant ERC-2016-STG-EvoluTEM-715502 and ERC-2020-COG-QTWIST-101001515). We also thank the Engineering and Physical Sciences Research Council (EPSRC) for funding under grants EP/Y024303, EP/S021531/1, EP/M010619/1, EP/V007033/1, EP/S030719/1, EP/V001914/1, EP/V036343/1 and EP/P009050/1, access to ARCHER2 supercomputer through the Materials Chemistry Consortium (EP/X035859), and also for the EPSRC Centre for Doctoral Training (CDT) Graphene-NOWNANO. TEM access was supported by the Henry Royce Institute for Advanced Materials, funded through EPSRC grants EP/R00661X/1, EP/S019367/1, EP/P025021/1, and EP/P025498/1. RVG. and VIF. acknowledge funding from the European Quantum Flagship Project 2DSIPC (no. 820378). We thank Diamond Light Source for access and support in use of the electron Physical Science Imaging Centre (Instrument E02 and proposal numbers MG33252 and MG35552) that contributed to the results presented here. SJH acknowledges funding from BP through the BP-International Centre for Advanced Materials. BD, SP, NFD and GJH would like to thank the Max Planck Centre on the Fundamentals of Heterogeneous Catalysis (FUNCAT) for funding. AJL acknowledges funding by the UKRI Future Leaders Fellowship program (MR/T018372/1, MR/Y034279/1).

**Author contributions:**

Conceptualization: SSA, NC, RG, SJH

Methodology: SSA, NC, WW

Investigation: SSA, NC, WW, RC, WT, DGH, JM, BD, SP, NFD, RZ, ML, GT, JH, HDL, JP, JS, EC, AC, DL, NM, AS

Formal Analysis: SSA, NC

Funding Acquisition: CSA, MD, AJL, VF, GJH, RG, SJH

Supervision: CSA, MD, AJL, VF, GJH, RG, SJH

Writing: SSA, NC, RG, SJH

**Competing interests:** Authors declare that they have no competing interests.

References and Notes:

1. A. Wang, J. Li, T. Zhang, Heterogeneous single-atom catalysis. *Nat. Rev. Chem.* **2**, 65–81 (2018).

2. X.-F. Yang, A. Wang, B. Qiao, J. Li, J. Liu, T. Zhang, Single-Atom Catalysts: A New Frontier in Heterogeneous Catalysis. *Acc. Chem. Res.* **46**, 1740–1748 (2013).

3. L. Liu, A. Corma, Metal Catalysts for Heterogeneous Catalysis: From Single Atoms to Nanoclusters and Nanoparticles. *Chem. Rev.* **118**, 4981–5079 (2018).

4. G. Malta, S. A. Kondrat, S. J. Freakley, C. J. Davies, L. Lu, S. Dawson, A. Thetford, E. K. Gibson, D. J. Morgan, W. Jones, P. P. Wells, P. Johnston, C. R. A. Catlow, C. J. Kiely, G. J. Hutchings, C. Richard, A. Catlow, C. J. Kiely, G. J. Hutchings, Identification of single-site gold catalysis in acetylene hydrochlorination. *Science* **355**, 1399–1403 (2017).

5. G. Malta, S. A. Kondrat, S. J. Freakley, C. J. Davies, S. Dawson, X. Liu, L. Lu, K. Dymkowski, F. Fernandez-Alonso, S. Mukhopadhyay, E. K. Gibson, P. P. Wells, S. F. Parker, C. J. Kiely, G. J. Hutchings, Deactivation of a Single-Site Gold-on-Carbon Acetylene Hydrochlorination Catalyst: An X-ray Absorption and Inelastic Neutron Scattering Study. *ACS Catal.* **8**, 8493–8505 (2018).

6. P. Johnston, N. Carthey, G. J. Hutchings, Discovery, Development, and Commercialization of Gold Catalysts for Acetylene Hydrochlorination. *J. Am. Chem. Soc.* **137**, 14548–14557 (2015).

7. X. Li, S. Mitchell, Y. Fang, J. Li, J. Perez-Ramirez, J. Lu, Advances in heterogeneous single-cluster catalysis. *Nat. Rev. Chem.* **7**, 754–767 (2023).

8. X. Sun, S. R. Dawson, T. E. Parmentier, G. Malta, T. E. Davies, Q. He, L. Lu, D. J. Morgan, N. Carthey, P. Johnston, S. A. Kondrat, S. J. Freakley, C. J. Kiely, G. J. Hutchings, Facile synthesis of precious-metal single-site catalysts using organic solvents. *Nat. Chem.* **12**, 560–567 (2020).

9. D. J. Kelly, M. Zhou, N. Clark, M. J. Hamer, E. A. Lewis, A. M. Rakowski, S. J. Haigh, R. V. Gorbachev, Nanometer Resolution Elemental Mapping in Graphene-Based TEM Liquid Cells. *Nano Lett.* **18**, 1168–1174 (2018).

10. N. Clark, D. J. Kelly, M. Zhou, Y.-C. Zou, C. W. Myung, D. G. Hopkinson, C. Schran, A. Michaelides, R. Gorbachev, S. J. Haigh, Tracking single adatoms in liquid in a Transmission Electron Microscope. *Nature* **609**, 942–947 (2022).

11. R. Frisenda, E. N. Navarro-Moratalla, P. Gant, D. Pérez De Lara, P. Jarillo-Herrero, R. V. Gorbachev, A. Castellanos-Gomez, D. Pé, D. Lara, P. Jarillo-Herrero, R. V. Gorbachev, A. Castellanos-Gomez, Recent progress in the assembly of nanodevices and van der Waals heterostructures by deterministic placement of 2D materials. *Chem. Soc. Rev.* **47**, 53–68 (2018).

12. M. F. Crook, I. A. Moreno-Hernandez, J. C. Ondry, J. Ciston, K. C. Bustillo, A. Vargas, A. P. Alivisatos, EELS Studies of Cerium Electrolyte Reveal Substantial Solute Concentration Effects in Graphene Liquid Cells. *J. Am. Chem. Soc.* **145**, 6648–6657 (2023).

13. W. Wang, N. Clark, M. Hamer, A. Carl, E. Tovari, S. Sullivan-Allsop, E. Tillotson, Y. Gao, H. De Latour, F. Selles, J. Howarth, E. G. Castanon, M. Zhou, H. Bai, X. Li, A. Weston, K. Watanabe, T. Taniguchi, C. Mattevi, T. H. Bointon, P. V. Wiper, A. J. Strudwick, L. A. Ponomarenko, A. V. Kretinin, S. J. Haigh, A. Summerfield, R. Gorbachev, Clean assembly of van der Waals heterostructures using silicon nitride membranes. *Nat. Electron.* **6**, 981–990 (2023).

14. M. Textor, N. De Jonge, Strategies for Preparing Graphene Liquid Cells for Transmission Electron Microscopy. *Nano Lett.* **18**, 3313–3321 (2018).

15. T. P. Hardcastle, C. R. Seabourne, R. Zan, R. M. D. Brydson, U. Bangert, Q. M. Ramasse, K. S. Novoselov, A. J. Scott, Mobile metal adatoms on single layer, bilayer, and trilayer graphene: An *ab initio* DFT study with van der Waals corrections correlated with electron microscopy data. *Phys. Rev. B* **87**, 195430 (2013).

16. H. E. Swanson, H. F. McMurdie, M. C. Morris, E. H. Evans, "Standard x-ray diffraction powder patterns" (NBS MONO 25-7, National Bureau of Standards, Gaithersburg, MD, 1969); https://doi.org/10.6028/NBS.MONO.25-7.

17. J.-P. Jalkanen, M. Halonen, D. Fernández-Torre, K. Laasonen, L. Halonen, A Computational Study of the Adsorption of Small Ag and Au Nanoclusters on Graphite. *J. Phys. Chem. A* **111**, 12317–12326 (2007).

18. J. Shan, C. Ye, Y. Jiang, M. Jaroniec, Y. Zheng, S.-Z. Qiao, Metal-metal interactions in correlated single-atom catalysts. *Sci. Adv.* **8**, eabo0762 (2022).

19. J. C. Crocker, D. G. Grier, Methods of Digital Video Microscopy for Colloidal Studies. *J. Colloid Interface Sci.* **179**, 298–310 (1996).

20. D. Allan, T. Caswell, N. Keim, C. van der Wel, R. Verweij, TrackPy, version 0.6.3 (2024); https://doi.org/10.5281/zenodo.11522100.

21. U. Patil, N. M. Caffrey, The role of solvent interfacial structural ordering in maintaining stable graphene dispersions. *2D Mater.* **11**, 015017 (2024).

22. S. Mitchell, F. Parés, D. Faust Akl, S. M. Collins, D. M. Kepaptsoglou, Q. M. Ramasse, D. Garcia-Gasulla, J. Pérez-Ramírez, N. López, Automated Image Analysis for Single-Atom Detection in Catalytic Materials by Transmission Electron Microscopy. *J. Am. Chem. Soc.* **144**, 8018–8029 (2022).

23. R. D. Deegan, O. Bakajin, T. F. Dupont, G. Huber, S. R. Nagel, T. A. Witten, Capillary flow as the cause of ring stains from dried liquid drops. *Nature* **389**, 827–829 (1997).

24. P. J. Yunker, T. Still, M. A. Lohr, A. G. Yodh, Suppression of the coffee-ring effect by shape-dependent capillary interactions. *Nature* **476**, 308–311 (2011).

25. T. Otanicar, J. Hoyt, M. Fahar, X. Jiang, R. A. Taylor, Experimental and numerical study on the optical properties and agglomeration of nanoparticle suspensions. *J. Nanoparticle Res.* **15**, 2039 (2013).